\documentclass[11pt]{amsart}
\usepackage{graphics}
\usepackage{hyperref}
\numberwithin{equation}{section} 
\def \dd{{\rm d}}

\begin{document}

\title[]{The topology of Schwarzschild's solution and the Kruskal metric}%

\author{Salvatore Antoci}%
\address{Dipartimento di Fisica ``A. Volta'' and INFM, Pavia, Italy}%
\email{Antoci@fisicavolta.unipv.it}%
\author{Dierck-Ekkehard  Liebscher}%
\address{Astrophysikalisches Institut Potsdam, Potsdam, Germany}%
\email{deliebscher@aip.de}%


\begin{abstract}
Kruskal's extension solves the problem of the arrow of time of the
``Schwarzschild solution'' through combining two Hilbert
manifolds by a singular coordinate transformation. We discuss the implications
for the singularity problem and the definition of the mass point.

The analogy set by Rindler between the Kruskal metric and the
Minkowski spacetime is investigated anew. The question is answered,
whether this analogy is limited to a similarity of the
chosen ``Bild\-r\"aume'', or can be given a deeper, intrinsic meaning.
The conclusion is reached by observing a usually neglected difference:
the left and right quadrants of Kruskal's metric are endowed with worldlines
of absolute rest, uniquely defined through each event by
the manifold itself, while such worldlines obviously do not exist
in the Minkowski spacetime.
\end{abstract}
\maketitle
\section{Introduction: Kruskal's extension of the Schwarzschild solution
and the arrow of time}
In general, a manifold cannot be covered by a single coordinate system.
An atlas of coordinate systems is required, and it reflects the global
properties of the manifold. An individual coordinate
system can be limited by its singularities, that may be seen as
singularities in the metric topology with respect to the coordinate-based
topology. Coordinates, however, do not matter. Any determination of position or time
is ruled by equations of motion or structure that
are invariant against substitutions of new coordinates for the old.
Hence, a singularity which is physically observable must have an expression
in invariant quantities.

In a Riemannian manifold, all tensors that can be constructed from the
metric tensor and its derivatives are the concomitants of the Riemann tensor.
For a general manifold, the invariants
of the Riemann tensor define singularities. When these invariants are
non-singular, the singularities that limit the viability of a coordinate
system are not observable -- this is the folklore. It holds, however,
only for a non-degenerate metric.
The ``Schwarzschild metric''\footnote[1]{Due to compelling historical reasons,
made clear \cite{AL2003} in an Editorial Note recently appeared in General
Relativity and Gravitation, and accompanying an English translation
of Schwarzschild's fundamental paper, we shall henceforth call Schwarzschild
solution, without quotation marks, the original ``Massenpunkt''
solution that Karl Schwarzschild published \cite{Schw16a} in 1916,
while the noun ``Schwarzschild solution'', or Hilbert solution, is reserved
to the metric manifold, inequivalent to Schwarzschild's original one,
later provided by Droste, Hilbert, Weyl \cite{Droste17, Hilbert17,Weyl17}
and attributed to Schwarzschild by these authors, as well as by nearly
all the subsequent writers of relativity.} shows
that in algebraically special metrics, where Killing vectors are
invariantly defined, singularities can now exist in the Killing
congruences also in the case of non-singular curvature invariants.
In the ``Schwarzschild metric'', it is the transition of
character of the Killing congruence from time-like to space-like that yields
a singularity in the acceleration: with $\xi^k = [0,0,0,1]$ and
$\xi_{i;k} + \xi_{k;i} = 0$,
we obtain
\begin{equation}\label{1.1}
\xi_{i;k}\xi^k + \frac{1}{2}(\xi_k\xi^k)_{,i} = 0\  ,
\end{equation}
and for $u_i = \frac{\xi_i}{\sqrt{g_{lm}\xi^l\xi^m}}$,
\begin{equation}\label{1.2}
u_{i;k}u^k + \frac{1}{2}(\ln \xi_k\xi^k)_{,i} = 0\  .
\end{equation}
The invariant square of the acceleration is found to be
\begin{equation}\label{1.3}
a_ia_kg^{ik} = \frac{1}{2}(\ln \xi_m\xi^m)_{,i}
\frac{1}{2}(\ln \xi_n\xi^n)_{,k} g^{ik} =
\frac{1}{4}g^\prime_{44}g^\prime_{44}g^{11}(g^{44})^2,
\end{equation}
where the prime denotes derivation with respect to $r$ in, say,
Hilbert's coordinates.
We find a singularity at the horizon not in curvature but in the
Killing congruence. It coincides with the change in character, and
it is measurable through the acceleration
necessary to keep a massive body stationary at a given position.

Beyond the Schwarzschild horizon, there is no observation for an observer
with a minimum distance from the horizon.
Any object falling onto the horizon will not reach it in finite observer time.
The proper time of the object, however, is finite, and the question of
extending the manifold beyond the horizon is posed.
This extension cannot be achieved when the Killing time is used as
time coordinate. The Killing time is not timelike at all ``inside''
the horizon: its vector is there spacelike.
This circumstance entails a confirmation of the argument, advanced
by Marcel Brillouin \cite{Brillouin23} already in 1923, and
reported in the Appendix, about a mismatch between the inner and the
outer problem. As remarked by Rindler \cite{Rindler90,
Rindler2001}, if the inner part of Hilbert's solution
is accepted as physical, due to the exchange of r\^ole
that the radial coordinate and the time coordinate
undergo at Schwarzschild's two-surface, there is no way of drawing
the arrows of time both inside and outside
in a way exempt from contradiction. A simple glance to
Fig. 1~(a),
\begin{figure}[ht]
\includegraphics{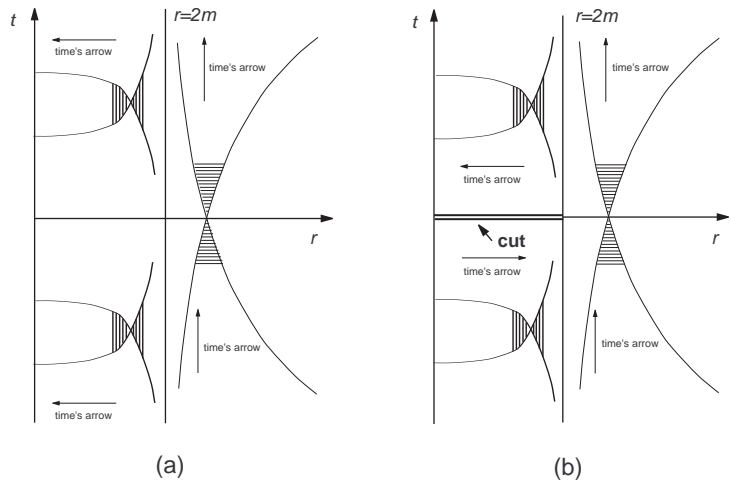}
\caption{ (a): $r$, $t$ representation of Hilbert's manifold. Light cones
and time's arrows are drawn, showing that the manifold cannot be endowed
with a consistent direction of time. (b): a cut is made in Hilbert's manifold,
thus allowing for a consistent choice of the time's arrows.}
\end{figure}
reported also by Rindler in his articles and books
\cite{Rindler90, Rindler2001}, suffices for gathering that we are
confronted with a contradiction. Moreover, it is not
a contradiction associated with a particular choice of the coordinates.
In fact, it is a flaw inherent in the geometric structure of the manifold
inadvertently chosen by Hilbert, a flaw that no one-to-one
coordinate transformation can remove,
no matter whether it is regular or not in the sense of Hilbert and Lichnerowicz
\cite{Hilbert17,Lichnerowicz55}.
This conclusion could have been sufficient for discarding
Hilbert's solution, with the inherent puzzle of the two
singularities at $r=0$ and at $r=2m$, and for adhering to the
solution that Karl Schwarzschild had originally proposed.

Coordinates do not matter -- locally. The postulate that they shall not matter globally
means calling manifolds of different (defined by the coordinates) topology equivalent.
This is not a necessary definition of equivalence classes. When one includes the
(defined by the coordinates) topology as characteristic of the manifold, only regular
(with regular and non-vanishing Jacobi determinant) coordinate transformations relate
equivalent manifolds. A part of the coordinate singularities become real, in particular
the zeros of the determinant of the metric tensor. These have been considered in the
early sixties \cite{THJ62,KLT63,KLT67}.

The ingenuity of the relativists, as is well known, took
instead a quite different route.
\section{Cutting and pasting two Hilbert manifolds}
The flaw of the Hilbert manifold
might be cured if it were possible to cut its inner region
as is shown in Fig. 1 (b), since then the two disjoint
inner parts can be endowed with opposite time arrows, thus
restoring the overall coherence in the direction of time.
Of course, after the cut we no longer have to do with
Hilbert's manifold, but with a new manifold, whose topological
properties and physical interpretation are different from those prevailing in
the previous one.
Moreover, the cut introduces an artificial border without
any physical raison d'\^etre, imposed merely by the need to solve
a contradiction. In order to heal the wound inferted to Hilbert's
manifold, by following Synge's original idea \cite{Synge50}, one decade later
Kruskal and Szekeres \cite{Kruskal60, Szekeres60} chose to effect
the following audacious act of surgery: paste together two Hilbert
manifolds, both affected by the same fatal illness, after suitably cutting them.
One is depicted in Fig. 1 (b); the second one is its mirror image
in the $r$, $t$ plane, left to the imagination of the reader. Then the upper
border of the cut in Fig. 1 (b) is sewn to the upper border
of the cut in the mirror image manifold, and likewise
for the lower borders. This way the cuts are sutured, and one
gladly recognises that, after the surgery, the circulation of the
arrows of time is exempt from contradiction. There is, however, a negative
consequence of the surgical act: it is the impairment
of the independence of the individual manifolds, with a future singularity
permanently sewn to its past counterpart. One may well ask: the Kruskal
extension, is it the solution for the field of a mass point?
It is, of course, when we consider only one of the regions outside the horizon.
It is difficult to swallow that the inner part should be considered as the
field of a particle: the extension converts the problem of the ``inner''
time arrow into the paradox of the dissolution of the particle into
an initial and a final singularity with vacuum in the time between.
\section{The analogy between the Kruskal metric and the Minkowski
spacetime}
The Kruskal manifold was soon noted for its considerable mathematical
beauty, but looked very remote, from a physical standpoint, from the
``Schwarzschild solution'' and from the problem that the latter
had tried to solve, after the original solution given by Schwarzschild
\cite{Schw16a} had been discarded. Since, due to the contradiction of
the arrows of time, Hilbert's solution was useless,
only the Kruskal solution seemed available for providing, within the theory
of general relativity, a metric that could be the counterpart of the field
of a point particle in Newtonian physics, but many a relativist, like Rindler,
felt necessary to show that the complexity of the Kruskal manifold
could be reduced
to something simpler, and more familiar to the physicists,
although of course not as simple and familiar as the gravitational
field of a point particle in Newton's theory.

First came an attempt \cite{Rindler65} at eliminating the strange
duplication present in the Kruskal-Szekeres metric, by identifying the
pair of events $(u,v,\vartheta,\varphi)$ and $(-u,-v,\vartheta,\varphi)$
in Kruskal's diagram, drawn in Fig. 2. This attempt was however unsuccessful,
due to the singularity that such an identification produces on the two-sphere
$u=v=0$, and to the ambiguity that this move
introduces in the direction of time. Then Rindler's attention was
drawn to the analogy that he had already brought to light five years earlier,
when dealing with the definition of motion with a constant acceleration in curved
spacetime \cite{Rindler60}. Let $u^i$ be the four-velocity
of a test particle in a pseudo Riemannian manifold, whose
affine connection is $\Gamma^i_{kl}$.
Then the absolute derivative of the four-velocity
\begin{equation}\label{3.1}
a^i\equiv\frac{\dd u^i}{\dd s}+\Gamma^i_{kl}u^ku^l
\end{equation}
defines the first curvature of the world-line of the particle,
{\em i.e.} its four-acceleration \cite{Synge60}.
Let the ``Schwarzschild metric'' be written in terms of
Hilbert's spherical polar coordinates $r$, $\vartheta$, $\varphi$, $t$, like
in footnote 4. It is an easy matter to verify \cite{Rindler60},
that, in keeping with Einstein's principle of equivalence,
\begin{figure}[ht]
\includegraphics{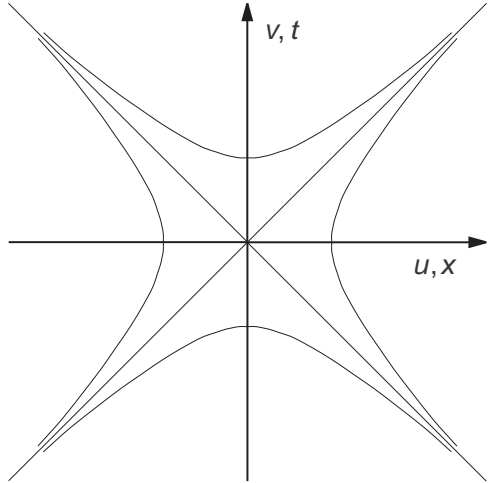}
\caption{In this figure the ``Bild\-raum'' for the Kruskal
manifold and a spacetime section of the Minkowski metric are
juxtaposed. The coordinate
axes are endowed both with Kruskal's $u$, $v$ coordinate labels,
and with the $x$, $t$ coordinates of the Minkowski reference system.
Therefore to each
point in the $u$, $v$ ``Bild\-raum'' is associated a two-sphere of
appropriate curvature radius, while to each point in the $x$, $t$
representative space  is associated a $y$, $z$ plane of the Minkowski
spacetime.}
\end{figure}
a test particle in ``Schwarzschild'' exterior metric, whose
spatial coordinates $r$, $\vartheta$, $\varphi$ are kept constant,
is subject to a radially directed four-acceleration, whose norm
\begin{equation}\label{3.2}
\alpha=(-a_ia^i)^{1/2}=\frac {1}{r^{3/2}(r-2m)^{1/2}},
\end{equation}
defined through equation (\ref{1.3}), is constant.
Let us now consider the $x$, $t$ plane of
Minkowski spacetime, drawn too in Fig. 2, and calculate the
four-acceleration of a test particle executing a hyperbolic
motion \cite{Rindler60} in, say, either the left or the right quadrant
of that plane. Its equation of motion shall read
\begin{equation}\label{3.3}
x^2-t^2=X^2,
\end{equation}
where the constant $X$ measures the minimum distance from the origin, attained
by the test particle when $t=0$, and the norm of the four-acceleration
of the test particle is now
\begin{equation}\label{3.4}
\alpha=\frac{1}{X}.
\end{equation}
Rindler noticed \cite{Rindler90, Rindler2001} that the particle
with constant spatial coordinates
in the ``Schwarzschild solution'', when viewed in the $u$, $v$
``Bild\-raum'', appears to execute a hyperbolic motion too, provided
that the Minkowski metric be substituted for the Kruskal metric
in either the left or the right quadrant of the Kruskal manifold.
On this initial basis he constructed an articulated argument
{\em ad analogiam} for gaining an understanding of the Kruskal
metric through the undisputable understanding of the Rindler
metric, which is merely a matter of special relativity.
\section{About using arguments {\em ad analogiam} in the theory
of general relativity}
A ``neo Cartesian'', meant in the jokeful sense that Synge once
attributed to the noun \cite{Synge66}, may well wonder what is
the r\^ole of arguments {\em ad analogiam} in the theory of general
relativity. One cannot in fact forget that, when Hilbert succeeded
\cite{Hilbert15} in formulating the field equations of the theory from
an action principle, he ended the Communication with which he
announced his achievement by expressing the hope that
\begin{quotation}
damit die M\"oglichkeit naher\"uckt, da{\ss} aus der Physik im
Prin\-zip eine Wissenschaft von der Art der Geometrie werde:
gewi{\ss} der herrlichste Ruhm der axiomatischen Methode, die hier
wie wir sehen die m\"achtigen Instrumente der Analysis, n\"amlich
Variationsrechnung und Invariantentheorie, in ihre Dienste nimmt.
\footnote[2]{An English translation:``hence the possibility gets close,
that from physics originate, in
principle, a science of the kind of geometry: certainly the most
splendid glory of the axiomatic method, that here, as we see,
takes to its service the powerful instruments of analysis, {\em i.e.}
calculus of variations and theory of invariants.''}
\end{quotation}
When Weyl and Levi-Civita obtained \cite{Weyl17, Levi-Civita19}
the reduction to quadratures for the solution of the vacuum field equations
of general relativity in the case of a static, axially symmetric manifold,
they did so by availing of the so called ``canonical
coordinates''.  Let $x^0=t$ be the time coordinate,
while $x^1=z$, $x^2=r$ are the coordinates in a meridian half-plane,
and $x^3=\varphi$ is the azimuth of such a half-plane; then the line
element of a static, axially symmetric field {\it in vacuo} can be written as:
\begin{equation}\label{4.1}
\dd s^2=e^{2\psi}dt^2-\dd\sigma^2,\;e^{2\psi}\dd\sigma^2
=r^2\dd\varphi^2+e^{2\gamma}(\dd r^2+\dd z^2);
\end{equation}
the two functions $\psi$ and $\gamma$ depend only on $z$ and $r$.
Remarkably enough, in the ``Bild\-raum'' introduced by Weyl $\psi$
fulfils the potential equation
\begin{equation}\label{4.2}
\Delta\psi=\frac{1}{r}\left\{\frac{\partial(r\psi_z)}
{\partial z}
+\frac{\partial(r\psi_r)}{\partial r}\right\}=0
\end{equation}
($\psi_z$, $\psi_r$ are the derivatives with respect to $z$ and to
$r$ respectively), while $\gamma$ is obtained by solving the system
\begin{equation}\label{4.3}
\gamma_z=2r\psi_z\psi_r,\;\gamma_r=r(\psi^2_r-\psi^2_z);
\end{equation}
due to the potential equation (\ref{4.2})
\begin{equation}\label{4.4}
\dd\gamma=2r\psi_z\psi_r\dd z+r(\psi^2_r-\psi^2_z)\dd r
\end{equation}
happens to be an exact differential.

The analogy of equation (\ref{4.2}) with the corresponding equation in
Newton's theory was indeed impressive, and mathematically helpful.
Neither Weyl nor Levi-Civita, however, thought of availing
of this analogy for gaining some insight in the physical meaning
of the solutions. They knew in fact that,
when Schwarzschild's solution \cite{Schw16a}
is rewritten by using Weyl's canonical coordinates,
the ``source'' for the ``Newtonian potential'' $\psi$,
that then happens to appear at the right-hand side of equation (\ref{4.2}),
looks like one segment of the $z$-axis covered by matter with
a constant linear mass density. But they also knew that this alluring analogy
produced in the ``Bild\-raum'' is deceitful.
As noticed by Weyl, the intrinsic form of the ``source'' is
a completely different one: the segment covered
by mass happens in fact to be Schwarzschild's two-surface.
\section{Intrinsic viewpoint. The absolute statics of general relativity}
Given the illusory character of the ``Bild\-raum'' analogy
in the case of the solutions of Weyl and Levi-Civita,
one shall verify whether the ``Bild\-raum'' analogy between
the Kruskal metric and Minkowski spacetime has some
intrinsic content too, that can be expressed in invariant way.
We find in Rindler \cite{Rindler90} the assertion that
``the Kruskal diagram possesses essentially the same invariance
properties as the Minkowski diagram''. We shall ask
whether this claim about a possibly deceitful correspondence between
``Bild\-raum'' invariance properties is really meaningful,
{\em i.e.} whether it can be turned into a like
claim about intrinsic invariance properties shared by the manifolds.

We have already noticed that while each $u$, $v$ point in Fig. 2
actually corresponds to a two-sphere, each $x$, $t$ point on the
same figure corresponds to a $y$, $z$ plane in Minkowski spacetime.
When $m\rightarrow 0$, both the original Hilbert manifold of Fig.
1~(a) and the manifold with a cut of Fig. 1~(b) tend to the
Minkowski spacetime. Kruskal's manifold is obtained
by sewing together, in the way described in Sec. 2, the cuts of two
manifolds like the one in Fig. 1~(b). Hence, in the limit
of vanishing mass, Kruskal's manifold happens to be constituted
by two Minkowski spacetimes, joined at the event $r=t=0$.
Therefore, already in the limit $m\rightarrow 0$, Kruskal's manifold
is topologically different from the Minkowski manifold.

An analogy of intrinsic character seems promised by the following
fact. A test particle executing hyperbolic motion in
either the left or the right quadrant of the $u$, $v$
``Bild\-raum'', and a test particle executing for good a hyperbolic
motion in Minkowski spacetime, in particular along the world lines
drawn in the $x$, $t$ plane of Fig. 2, both execute a motion whose
four-acceleration is constant in direction and in norm,
as shown by Eqs. (\ref{3.2}) and (\ref{3.4}) respectively.
Are these particular motions an intrinsic, common feature of both
the Kruskal and the Minkowski manifold?

In the Kruskal manifold, these motions occur in quadrants
that are static in character, if one accepts the usual
definition of ``static''. Let Greek indices run from 1 to 3.
According to the usual definition, a region of a manifold is
static if a coordinate system can be chosen for it, in
which the square of the interval can be written in
the form
\begin{equation}\label{5.1}
\dd s^2=g_{44}\dd t^2+g_{\mu\nu}\dd x^{\mu}\dd x^{\nu},
\end{equation}
and the nonvanishing components of the metric do not depend on
the timelike coordinate $t$.
According to this definition, both the left and right Kruskal
quadrants and the Minkowski spacetime are static; this fact
appears to strengthen the analogy between the  motions with constant
acceleration occurring in these manifolds.

The attribution of the adjective ``static'' to the Minkowski
manifold and to the Minkowski metric does not seem however
to be an entirely convincing one. In fact, the notion of staticity
is indisputably connected with the notion of rest. How is it possible
to call ``static'' the metric of the theory of special relativity, a theory that
denies intrinsic meaning to the very notion of rest? How is it
possible that the Minkowski metric remain static, as it does
according to our definition, after we have subjected it to an
arbitrary Lorentz transformation, {\em i.e.} a transformation that
entails relative, uniform motion? Let us start from the Minkowski
metric, given by $g_{ik}={\rm diag}(-1,-1,-1,1)$ with respect
to the Galilean coordinates $x$, $y$, $z$, $t$, and perform the
coordinate transformation
\begin{equation}\label{5.2}
x=X\cosh{T}, \ y=Y, \ z=Z, \ t=X\sinh{T}
\end{equation}
to new coordinates $X$, $Y$, $Z$, $T$.
We get the particular Rindler metric \cite{Levi-Civita18},
whose interval reads
\begin{equation}\label{5.3}
\dd s^2=-\dd X^2-\dd Y^2-\dd Z^2+X^2\dd T^2.
\end{equation}
How is it possible that this metric turn out to be in static form too,
despite the fact that the transformation
(\ref{5.2}) entails uniformly accelerated motion?
There is however another definition of static manifold and metric,
that, although equivalent to the one given previously, is
more helpful for solving our difficulty, because it is
expressed in intrinsic language. It says that a metric is static
if it allows for a timelike Killing field $\xi_{i}$ that is also
hypersurface orthogonal, {\em i.e.} fulfils both the equations
\begin{equation}\label{5.4}
\xi_{i;k}+\xi_{k;i}=0,
\end{equation}
and
\begin{equation}\label{5.5}
\xi_{[i}\xi_{k,l]}=0.
\end{equation}
Both the Kruskal metric in the left and right quadrants
and the Minkowski metric possess timelike Killing fields that
obey both equations (\ref{5.4}) and (\ref{5.5}). There is however a
fundamental difference between the two cases, whose importance
has been somewhat neglected until now. In the case of the Minkowski
metric, at a given event there is an infinity of timelike
Killing vectors that fulfil both equations (\ref{5.4}) and (\ref{5.5}),
and this circumstance answers our previous questions, since it
says in intrinsic language that there is no privileged time axis,
hence no privileged rest frame in the manifold of special
relativity.

On the contrary, in the case of the left and right quadrants
of the Kruskal metric, when $m>0$ the timelike vector
that obeys both equations (\ref{5.4}) and (\ref{5.5}) is unique:
at any event in the left and right quadrants of the Kruskal
manifold the metric itself provides a unique, absolute
time direction; this circumstance allows one to draw through any event
a unique worldline of absolute rest, that is intrinsic
to the manifold. Hence, a worldline in the Hilbert manifold
for which $r>2m$, $\vartheta$ and $\varphi$ have constant values
is an intrinsic worldline of absolute rest.

From this result it follows that the norm of the
four-acceleration $a^i$ of a test particle
on one of these worldlines, given by (\ref{3.2}), is an invariant
and intrinsic property of the Hilbert and of the Kruskal manifolds.
In fact, this scalar is a unique outcome of equations
(\ref{5.4}), (\ref{5.5}) and (\ref{3.2}). We calculate
it by availing only of the metric,
without making any arbitrary choice. This scalar happens
to diverge when $r\rightarrow 2m$, as shown by equation (\ref{3.2}).
Due to the way kept in calculating the norm (\ref{3.2}), its divergence for
$r\rightarrow 2m$ is as intrinsic to the Hilbert and Kruskal manifolds
\cite{AL2001, ALM2001} as the divergence, for $r\rightarrow 0$, of
the scalars built with the Riemann tensor and with its covariant derivatives.

Also the norm (\ref{3.4}) of the four-acceleration of a test particle
executing a hyperbolic motion in the Minkowski manifold, {\em i.e.} staying
on a worldline  with constant values of $X$, $Y$, $Z$ in
the particular form (\ref{5.3}) of Rindler's metric, diverges
when $X\rightarrow 0$. This divergent behaviour is however
completely different in nature from the one occurring in the
Kruskal and Hilbert manifolds for $r\rightarrow 2m$. In fact, the
scalar (\ref{3.4}) is by no means uniquely dictated by the metric:
in the Minkowski manifold there is an infinity of worldlines of
hyperbolic motion through a given event; choosing one of them is
completely arbitrary, and such is also, from an intrinsic viewpoint,
the definition of the divergent scalar.

\section{Appendix: Historical note}
When, by further elaborating the  geometric interpretation
of the gravitational field of a particle envisaged by Synge
\cite{Synge50} in 1950, ten years later Kruskal \cite{Kruskal60}
and Szekeres \cite{Szekeres60} succeeded in convincing an
influential minority within the relativists that a major
breakthrough had been achieved in the mathematical understanding
of the ``Schwarzschild solution'', a major problem
became soon apparent too: convincing
that community that a major breakthrough had been achieved also
in the physical understanding. This was by no means an easy task:
the possibility that the Kruskal extension might represent
such an achievement was {\em e.g.} simply dismissed in
1962 by Dirac \cite{Dirac62} with a polite touch of nonchalant irony.
He noticed:
\begin{quotation}
The mathematicians can go beyond
this Schwarzschild radius, and get inside, but I would maintain
that this inside region is not physical space, because to send a
signal inside and get it out again would take an infinite time, so
I feel that the space inside the Schwarzschild radius must belong
to a different universe and should not be taken into account in
any physical theory.
\end{quotation}
Dirac's opinion had already been reached \cite{Brillouin23} four decades earlier
by Marcel Brillouin, although by availing of a different argument. In
1923, after pondering for months the problem of
the ``catastrophe Hadamard'' occurring at Schwarzschild's singular surface,
that had been raised during the discussions between Einstein and
the French scientists gathered at the ``Coll\`ege de France'' during the
Easter of 1922, Brillouin eventually wrote \cite{Brillouin23}:
\begin{quotation}
On peut se demander si cette singularit\'e
\footnote[3]{In Brillouin's paper, the metric
\begin{equation}\\\nonumber
\dd s^2=\gamma c^2\dd\tau^2-\frac{1}{\gamma}\dd
R^2-(R+2m)^2(\dd\theta^2+\sin^2\theta\dd\varphi^2), \ \
\gamma=\frac{R}{R+2m}
\end{equation}
is given. $c$ is the velocity of light,
$\theta$ and $\varphi$ are the usual spherical polar angles,
$R$ is the radial coordinate. Brillouin's pondered choice for the limit
values of $R$ ($0<R<\infty$) puts his  metric in one to one correspondence with
Schwarzschild's original metric \cite{Schw16a}.} {\em limite} l'Univers,
et s'il faut s'arr\`eter a $R=0$; ou si, au contraire, elle {\em traverse}
seulement l'Univers, qui continuerait au del\`a, pour $R<0$. Dans
les discussions, en particulier,
dans celles da P\`aques 1922, au Coll\`ege de France, on a
g\'en\'eralement parl\'e comme si $R=0$ caract\'erisait une region
{\em catastrophique} qu'il faut depasser pour arriver jusq'\`a la
v\'eritable limite singuli\`ere, atteinte seulement pour $\gamma$
infini,
avec $R=-2m$. {\em A mon avis, c'est la premi\`ere singularit\'e,
atteinte $R=0$, $\gamma=0$ $(m>0)$, qui limite l'Univers} et qu'on
ne doit pas d\'epasser. [{\em C. R.}, {\bf t}. 175 (27 nov.
1922)].

La raison en est p\'eremptoire, quoiqu'on ait n\'eglig\'e de la
mettre en \'evidence jusqu'\`a pr\'esent: {\em Pour $R<0$,
$\gamma<0$, le $\dd s^2$ ne correspond plus de tout au probl\`eme
qu'on voulait traiter.}
\end{quotation}
With hindsight, it must be recognised that Brillouin had given one
good reason for discarding Hilbert's form of the spherically symmetric
manifold of a massive particle \footnote[4]{It was Leonard Abrams, in a keen paper
\cite{Abrams89} written in 1989, who first noticed that the apparent
necessity of the choice of the manifold done by Hilbert when he wrote
\cite{Hilbert17} the spherically symmetric solution in the form:
\begin{equation}\nonumber
ds^2=\left(1-\frac{2m}{r}\right)dt^2
-\left(1-\frac{2m}{r}\right)^{-1}{dr^2}
-r^2(d\vartheta^2+\sin^2\vartheta d\varphi^2), \ \ 0<r<\infty,
\end{equation}
and handed it down to the posterity as the unique
``Schwarzschild solution'', is just due to an arbitrary restriction inadvertently
imposed by Hilbert in his calculation.}, with its troublesome inner region,
$0<r<2m$, and for adopting instead the manifold that Schwarzschild had
deliberately decided to choose \cite{Schw16a}, when he first
solved the ``Massenpunkt'' problem. The issue, whether Schwarz\-schild's
two-surface is intrinsically singular or not, and whether transformations of
coordinates that cancel the ``catastrophe Hadamard" are allowed,
despite the fact that they infringe the rules for the regularity of
coordinate transformations laid down by Hilbert \cite{Hilbert17} already
in 1917, and sharpened by Lichnerowicz \cite{Lichnerowicz55} in his book
of 1955, is with us since the very beginning of general
relativity. It has been widely debated with opposite outcomes, if it is true that
during forty years since its discovery, the overwhelming majority of relativists
was convinced that the inner region of Hilbert's solution was
unreachable from the outside, while during the subsequent
four decades an equally overwhelming
majority harbored the opposite conviction.

Brillouin's argument, however, does not rely at all on the singular
character of Schwarzschild's two-surface. It simply says \cite{Brillouin23} that,
because in the inner region of Hilbert's metric the radial coordinate
and the time coordinate exchange their r\^oles as spatial
and temporal coordinate respectively, in that region one is no
longer considering the problem that one had set out to solve.
\newpage
\bibliographystyle{amsplain}

\end{document}